# Spin-exchange-induced dimerization of an atomic 1-D system

Nader Zaki[1*], Chris A. Marianetti[1], Danda P. Acharya[2], Percy Zahl[2], Peter Sutter[2], Junichi Okamoto[1], Peter D. Johnson[3], Andrew J. Millis[1], Richard M. Osgood[1*]

[1]Columbia University, New York, NY 10027, USA.

[2]Center for Functional Nanomaterials, Brookhaven National Laboratory, Upton, NY 11973, USA.

[3]Condensed Matter Physics & Materials Science, Brookhaven National Laboratory, Upton, NY 11973, USA.

**Using low-temperature scanning tunneling microscopy, we demonstrate an unambiguous 1-D system that surprisingly undergoes a CDW instability on a *metallic* substrate. Our ability to directly and quantitatively measure the structural distortion of this system provides an accurate reference for comparison with first principles theory. In comparison to previously proposed physical mechanisms, we attribute this particular 1-D CDW instability to a ferromagnetic state. We show that though the linear arrayed dimers are not electronically isolated, they are magnetically independent, and hence can potentially serve as a binary spin-memory system.**

As has been shown with 0-D quantum dots and 2-D single-layer atomic sheets, 1-D atomic systems are expected to exhibit novel quantum mechanical phenomena due to angstrom-scale confinement[1,2]. The realization of true self-assembled monatomic 1-D systems which exhibit rich phenomena beyond quantum-well-like behavior[3], however, has been rare[4,5]. Most attempts have consisted of wire growth on semiconducting surfaces[6-11], in order to reduce substrate interaction, and have often observed a charge-density-wave (CDW) instability[2] commensurate with a metal-to-insulator transition. However, these systems suffer from the inability to directly determine the crystallographic structure and element constituency of the wires and the modified substrate surface[2,11]. Furthermore, there remains disagreement on the physical cause of the distortion, whether it be Peierls, Fermi nesting, or phonon interaction[2].

One prospectively important point is that most existing reports[2] of monatomic wire structural instabilities involve systems composed of heavy metal elements such as Au and In with completely filled *d*-shells and partially filled *s*- and *p*-orbitals. These limited examples raise questions as to whether density wave instabilities generically occur, in particular for light elements with partially filled *d*-orbitals, and if they can occur on substrates besides semiconducting surfaces.

Here we present low temperature scanning tunneling microscopy (LT-STM) observations of a dimerization (bond-centered density wave with a wavevector of $k = \frac{\pi}{a}$) instability of a Co atomic-wire system self-assembled on a vicinal Cu(111) substrate. Using *ab-initio* theoretical calculations we show that the partially filled *d*-shells in fact drive the instability. Further, in contrast to the previously studied systems, the instability in Co atomic wires is a consequence of strong local correlations on the Co atoms. The strong local correlations result in the *d*-shell of each Co ion being in a high spin configuration; the combination of locally maximal spin and partial orbital filling leads to ferromagnetic correlations, which enhance the dimerization instability.

The self-assembly based growth procedure, described elsewhere[12, 13], maximized Co-wire nucleation at the Cu step edges and minimized Cu-terrace substitution by Co atoms as well as Co island formation, as evidenced by the sparse presence of Co atoms in/on the step terrace (Fig. S1-S3). Unlike the case of self-assembled chains on semiconducting substrates[2, 11], the location of the atomic chains in this bi-metallic system is unambiguous.

An example of a LT-STM measurement of the self-assembled Co chains is shown in Fig. 1e. The measurement shows two Cu step edges, visible as a corrugation of the background topography, along which Co atoms (visible as features above the background level) have arranged themselves to form a 1-D chain. The chains do not consist of equally spaced Co atoms. A lateral distortion is clearly evident, indicating that the chain is dimerized. The typical measured 1-D unit cell width of 5.1Å matches well with twice the Cu-Cu atom spacing of the ideal Cu(111) substrate ($2 \times 2.56$Å).

The measured Co-Co bond length of 2.0($\pm$0.1)Å (see supporting discussion, Fig. S4) is noticeably shorter than the Cu-Cu atom spacing of 2.56Å, based on measurements over a set of chains. By comparison, the Co-Co atom distance in self-assembled Co triangle islands on Cu(111) ranges from 2.50Å to 2.56Å[14], which is comparable to that found in bulk Co. It is surprising that the Co chain would exhibit a structural distortion; given that the atom spacing of bulk Co is similar to bulk Cu (a difference of only ~2%), an ideal uniform Co atom spacing would be expected. It therefore appears that the bond length distortion in this case is due to the one-dimensional geometry along the Cu step edge with its anisotropic environment.

Our experimental observations provide a crucial test of first-principles methods, given that the Co chain may have a non-trivial degree of electronic correlations which is challenging for DFT calculations to describe. Therefore, we set out not to precisely model our experimental setup of a Co chain on a Cu step, but instead to understand how and why an isolated Co chain restricted to distortions in one spatial dimension dimerizes. The effect of the step may be subsequently deduced. *Ab initio* and density functional theory (DFT) calculations have previously been performed for free[15] and surface supported finite[16] and infinite atomic chains[17] with results ranging from non-dimerized to zig-zag to anisotropically strained. However, no clear physical mechanism has been deduced or put forward. Here we present DFT calculations, based on several different functionals (Fig. S5, S6)[12], which shed light on the physics underlying our experimental observations. Specifically, the energy of an infinite length 1-D periodic system consisting of two Co atoms per unit cell was studied under the constraint that the period of the system matched twice the Cu atom-atom spacing (2 × 2.56Å) and that the Co atoms are allowed to move only along the wire direction. The physical configuration was identified as the one which minimized the energy.

In the theoretical model there are two Co-Co bond lengths. Fig. 2a shows the dependence of the energy on the length of the shorter, i.e. nearest-neighbor, Co-Co bond (measured relative to the mean Co-Co distance). A clear energy minimum is visible at $d_{short} = 0.794\ d_{avg} = 2.03$Å (implying $d_{long} = 3.08$Å). A key result of the DFT calculation is that the Co *d*-shell on each site is essentially fully spin polarized, having maximal spin polarization for given d occupancy. Different orientations of the Co spin were investigated (Fig. 2a shows as an example the energy of the two sublattice antiferromagnet); the ground state was found to be ferromagnetic. Furthermore, the ferromagnetic phase favors a structural distortion while the antiferromagnetic phase does not. These finding suggest that the dimerization instability is driven by the energetics of electron transfer between *d*-orbitals subject to a ground state of maximal spin. Because the *d*-orbitals are partially occupied, transfer is optimized in a ferromagnetic state, while the high spin state means that electron transfer is essentially forbidden in the antiferromagnetic state. These considerations suggest that the spin polarized *d*-orbitals play a key role in the dimerization phenomenon.

To further investigate the relevance of the *d*- orbitals to the dimerization we compare in Fig. 2b the dimerization energetics of stretched wires of Co (partially filled *d* shell; DFT predicts ferromagnetic ground state) and Cu (fully filled *d* shell; DFT predicts paramagnetic ground state) wires. On general grounds we expect that a physical 1-D system that is stretched to have a mean bond length sufficiently far from its ideal bond-length spacing will undergo a distortion. However, both the extent of the strain required for dimerization and the amplitude of the distortion will depend on the physics responsible for the instability. In the two panels of Fig. 2b we plot the energy (relative the energy of the undimerized state) vs degree of dimerization, for different amounts of strain relative to the bond length which minimizes the DFT energy for the undimerized wire. We see that in the Co system the dimerization becomes favored at a much lower strain than in the Cu system, and the energy gain from dimerization is much greater for equal amounts of strain. We also compare the optimal length of the short bond (indicated by red vertical line) to the optimal spacing of the transition metal dimer (left grey dashed line) and to the optimal atom-atom distance in the unstrained wire (right grey line). For a strained Co wire, the nearest-neighbor bond length generally lies between the optimal dimer bond length and the optimal non-distorted wire bond length, indicating that dimerization is truly favored. By contrast, in the case of a strained Cu wire, the nearest-neighbor bond length is generally greater than, but close to, the optimal non-distorted wire bond length (at least for strains up to ~22%).

The comparison of Co and Cu highlights the difference in tendency to dimerization in a partially filled $d$-orbital derived band 1-D system vs a partially filled sp-orbital derived band 1-D system. However, it does not address the physical phenomenon that favors a stronger distortion in the Co wire system. The absence of the dimerization in the antiferromagnetic state indicates that the dimerization is connected to hopping of $d$-electrons (suppressed in the AF state by the condition that each Co ion is in a high spin configuration). The total $d$ occupancy is approximately $d^8$. Choosing an angular momentum quantization axis ($z$-direction) parallel to the chain direction we note that the $3d_{xy}$ and $3d_{x^2-y^2}$ orbital do not hybridize much along the chain and instead act as local moments (Fig. S7). The physics is driven by the $3d_{xz}$, $3d_{yz}$ and $3d_{z^2}$ orbitals. In the spin polarized state, the majority orbitals are filled while the minority orbitals are nearly ½ filled. In this circumstance, a dimerization instability leads to a large energy gain arising because in each orbital there is one minority spin electron per pair of atoms; this electron forms a strong bond in the dimerized state, with the antibonding orbital completely empty. This can be viewed as a Peierls distortion. Note that if the ground state was not high spin, the band energetics would be less favorable to dimerization because one would have partially, not fully occupied bonds. The dimerization instability is also favored by the relatively localized nature of the $d$-electrons which, in contrast to the more spatially extended $s$-$p$ electrons, have predominately a nearest neighbor hopping, and which moreover rises rapidly as the inter-atom distance is decreased.

The above physical explanation is only reasonable if the hybridization of the Co $d$-states to the electrons in the vicinal Cu substrate is relatively weak. This hypothesis is consistent with the observation of weak indirect spin exchange interaction for Co dimers on Au(111) and Cu(100) substrates[18, 19]. These studies found that though there exists a strong indirect exchange for a Co monomer by way of the Kondo effect (indicating the presence of a moment, i.e. a locally high spin configuration, on the Co site), a Kondo signal was lacking for a Co dimer that was fabricated by STM atom-tip manipulation. The lack of a Kondo signal for the dimer is naturally understood in terms of the non-negligible *d-d* electron transfer suggested here.

An important consequence of the distortion is a strongly decreased electron transfer between the 2-Co-atom unit cells, implying also that the coupling of spin between dimers becomes negligible. In order to quantify this, we perform a cluster expansion of the total energy in terms of the spin cluster functions. Given a lattice model with a binary site variable (i.e. up/down spin), one can perform a power series expansion of any average lattice observable in terms of the correlation functions of the site variables. In spin systems, one can often obtain a highly accurate expansion using only pair terms over a short range. We find that one can accurately represent the energetics using only neighbor pair terms, as defined in the following equation:

$$H = E_0 + \sum_i J_1 s_i \cdot s_{i+1} + J_2 s_{i+1} \cdot s_{i+2}$$

where $E_0$ is the non-magnetic energy contribution, $s$ is $\pm 1$, and $J_1/J_2$ are the neighbor magnetic pair interactions (Fig 3a, Fig. S8a)[12]. The goal of using this basic model is to show the change in these parameters as a function of distortion. In Fig. 3b, a plot of the parameters is shown (Fig. S8). Note that in the undistorted wire the magnetic interaction constants are equal by symmetry, with the negative sign arising because DFT favors a ferromagnetic state. As the system is distorted, $J_1$ increases rapidly in magnitude whereas $J_2$ goes quickly to zero. We also present the sum of the magnetic interaction parameters ($J_1+J_2$), which gives the total magnetic contribution to the energy. The monotonic decrease of the sum ($J_1$+$J_2$) as distortion is increased also shows that the ferromagnetic state strongly favors the distortion, as expected if the driving force is electron transfer between high-spin configuration $d$ states. Note that $E_0$ is monotonically increasing; this illustrates that non-magnetic terms do not play a role in the dimerization.

The striking difference in the variation of the magnetic interaction parameters, $J_1$ and $J_2$, under dimerization suggests that the spin chain may provide an interesting realization of a memory device. In the dimerized state, $J_2$ is negligible compared to the quite large magnetic coupling, $J_1$, between nearest-neighbors. A consequence of this result is that while each dimer is itself in a high spin state, the spins of neighboring dimers may take arbitrary orientations with negligible energy penalty. Hence, a dimerized Co spin-chain can potentially behave as a linear array of spin memory bits. Binary memory requires a bistability, in other words an easy axis for the magnetization of a dimer. We expect that this is provided by the Cu step edge, as was shown for a Co/Pt(997) system[20]. A somewhat similar system, but based on antiferromagnetic switching, has been recently realized using Fe on $Cu_2N$[21]. Finally, we note that even though the Co chain unit cells may be "spin-isolated", they are not electronically disjoint; an energy band diagram of the Co chain reveals band crossings (predominately of $d_{z^2}$ and $s$ character) at the Fermi energy (Fig. S7), suggesting a possibility for manipulating the spin states via appropriately applied currents[22].

A field of much current interest is that of suspended atomic chains formed by break-junctions[23]. While it has been found that suspended non-magnetic wires can be formed by this method, forming suspended magnetic wires has not yet been successful[24]. The reasoning for this was reported to be softening of the binding energy of the atomic chain due to magnetism[24]. While the process of forming a suspended chain is complicated[25], the results reported here suggest that forming a suspended Co atomic chain is difficult, in part, due to the tendency for the chain to dimerize. The long bond would be weakened due to the extent of the dimerization occurring during stretching of the chain (Fig. 2b).

One expects that the lattice distortion is a low temperature phenomenon, occurring only below a critical temperature, as observed for wires on semiconducting surfaces. The bi-metallic Co/Cu(775) system investigated here undergoes a phase transition at elevated temperatures. At a temperature of 91K, the Co distortion is non-uniform along a chain, varying from 0.6Å to 0.0Å. The Co chains also show a tip-bias dependency; at low tip bias, single Co atoms are easily resolved, while at higher tip bias, a ×2 periodicity is more prevalent (Fig. S9, S10). At a slightly lower temperature of 81(±4) K, however, some chains appear exactly like those measured at 5K, i.e. possess a dimerization instability that is independent of tip-bias (Fig. S11). These observations indicate a coexistence of two different phases, lending speculatively to the assignment of this system change as a $1^{st}$-order phase transition with a critical temperature in the vicinity of 100K.

Thus both our experimental and theoretical results show a Co-atomic wire on stepped Cu(111) behaves as a 1-D atomic system with a low-temperature spin-exchange-induced dimerization instability. This work raises the question as to whether other light partially filled *d*-orbital 1-D systems will exhibit a similar instability once realized in experiment, and their possible technological applications.

This work was supported by the Department of Energy Contract No. DE-FG 02-04-ER-46157. Work at Brookhaven National Laboratory was supported by the Department of Energy under Contract No. DE-AC02- 98CH10886. We thank Mark Hybertsen for discussions, suggestions, and help in performing the DFT calculations. We thank James Davenport and Abhay Pasupathy for helpful discussions.

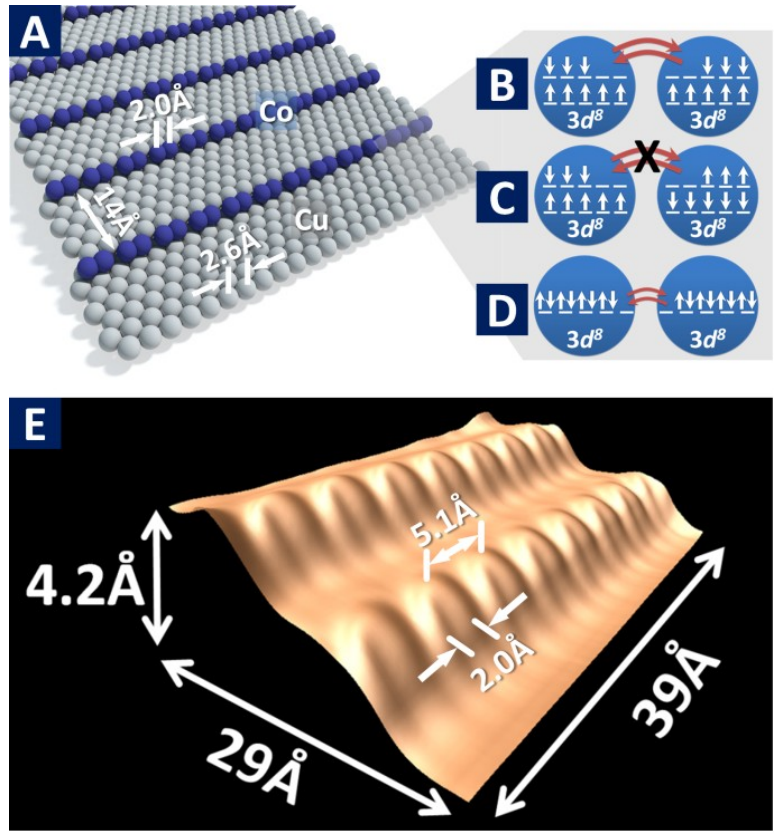

**Figure 1 Self-Assembled Co atomic chain system.** (a) Illustration of dimerized Co atomic chains on vicinal (8.5º miscut) Cu(111). (b)-(d) Illustration of high-spin ferromagnetic, high-spin anti-ferromagnetic, and zero-spin electron configurations. Coupling is strongest for the high spin ferromagnetic phase, weaker for the zero-spin phase, and blocked for the anti-ferromagnetic phase. (e) Perspective view of a STM topography of two self-assembled Co wires at adjacent Cu step edges. The vertical scale has been magnified to accentuate the appearance of the Co wires. The Co atoms constituting these wires have undergone a 1-D structural distortion, leading to the appearance of a single peak near the Cu step edge. However, the underlying Co atoms constituting each single peak are resolved farther away from the Cu step edge due to the decreased contribution of the Cu local density of states to the tunneling current. Constant current tunneling parameters: $V_{bias}$ = +0.742V at 9.4nA.

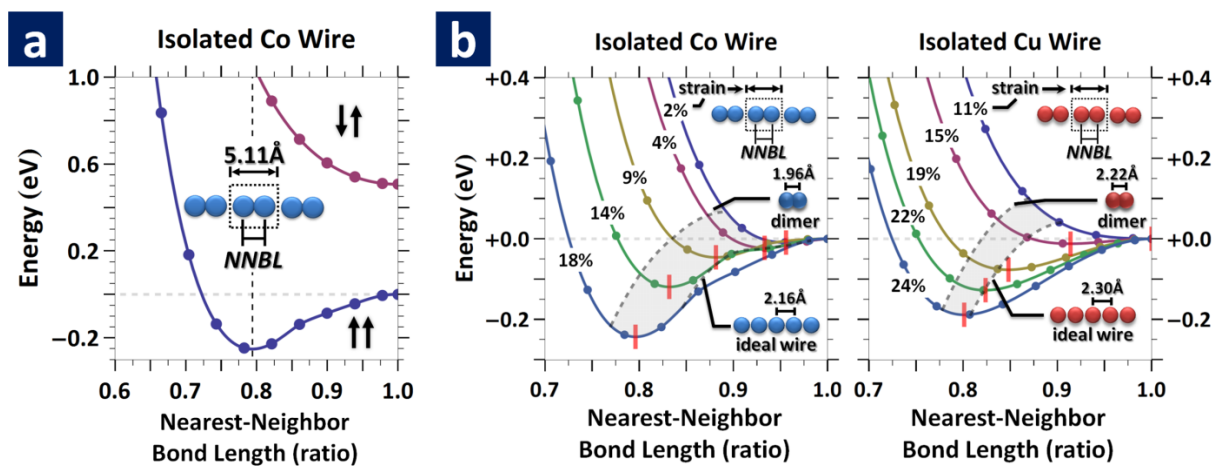

**Figure 2 DFT energy-phase diagrams for atomic wires.** Points denote actual calculated energies while smooth lines in between are 3$^{rd}$-order interpolations. Total energies were calculated for a two-atom unit cell using DFT and the GGA functional (Fig. S5, S6). The lack of smoothness in some regions is due to varying degrees of orbital polarization (see supporting discussion). (a) Energy-phase diagram for Co atomic wire. The energies of the Co wires have been offset with respect to the *ferromagnetic* non-distorted case (NNBL=1.0); the horizontal dashed grey line denotes the reference. The nearest neighbor bond length is given as a ratio of the bulk Cu atom spacing (2.5561Å). (b) Energy of Co$_2$ and Cu$_2$ isolated wires measured relative to the energy of the undimerized wire and plotted against degree of dimerization (parametrized as ratio of short bond length to average bond length). Different curves indicate different strains (i.e. different unit cell lengths) relative to the unit cell length that minimizes the DFT energy of the wire. The energy of the Co wire has been computed for a ferromagnetic state; the Cu wire was computed for a non-spin polarized state (spin-polarized calculations converged to a zero-spin state). The left hand dashed grey line denotes the optimal bond length of the dimer (measured in units of one half of the mean unit cell length of the strained wires). The right hand grey line indicates the interatomic distance for the optimal non-distorted wire. The red vertical line segments mark the optimal nearest-neighbor bond length for each respective strained wire system.

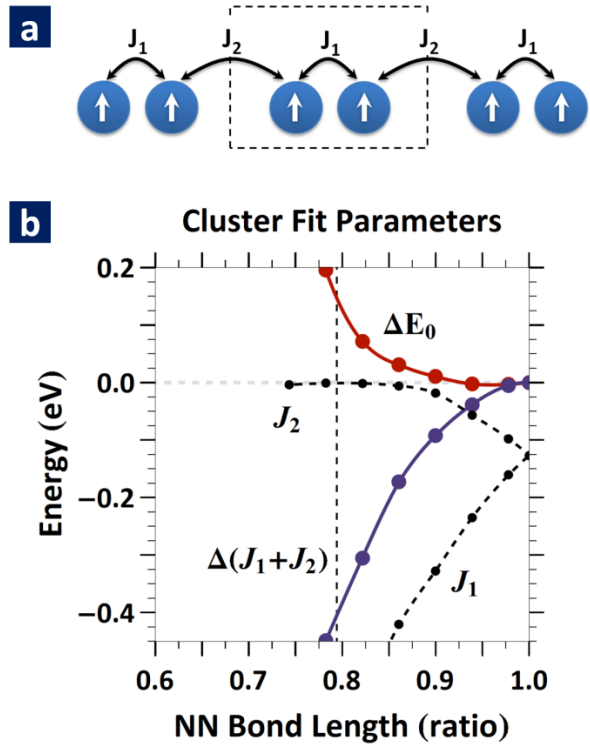

**Figure 3 Fitting to a cluster expansion model. a**, Cluster expansion model illustration of the Co 2-atom unit cell. $J_1$ denotes the nearest-neighbor or *intra*-pair interaction while $J_2$ denotes the *inter*-pair interaction. The unit cell is delineated by the dashed rectangle outline. **b**, Cluster expansion model parameter fits for the Co atom wire for different amounts of distortion. $E_0$ is the non-magnetic energy contribution. The vertical dashed line denotes the optimal Co wire distortion. Note that $J_2$ is quite negligible at the optimal distortion length. Energies were computed using spin polarized GGA (Fig. S8).

# Supplementary Information

## Spin-exchange-induced dimerization of an atomic 1-D system

N. Zaki, C. A. Marianetti, D. P. Acharya, P. Zahl, P. Sutter, J. Okamoto, P. D. Johnson, A. J. Millis, R. M. Osgood

R.M.O. (osgood@columbia.edu); N.Z. (nz2137@columbia.edu)

**Supporting Information Components:**

- Methods
- Supporting Discussion
- References (26-33)
- Author Contributions
- Figures S1-S11

**Methods**

Self-assembly of Co chains on Cu(775) was carried out in a UHV preparation chamber with a base pressure in the low $10^{-10}$ Torr range. Subsequent to cleaning the vicinal substrate by several sputter-$D_2$-anneal cycles, deposition of Co atoms by e-beam heating of a Co rod was performed at 130K using a slow deposition rate of 0.01ML/min. The deposition was followed by a brief increase in the sample temperature to ~RT to allow the Co atoms to migrate to the step edges and anneal any defects. Subsequently the sample was *in-situ* transferred to the STM mounting and then cooled to the experimental temperature (typically 5K). The tunneling microscopy measurements used a chemically etched tungsten tip. All measurements were performed at the Center for Functional Nanomaterials (CFN) LT-STM facility at Brookhaven National Lab. STM topography data was smoothed using a Gaussian filter in order to remove noise derived artifacts. Some of the 3-D topography images were generated using WSXM[26].

It is possible to resolve the Co chains more clearly from the STM topographical measurements by use of a technique akin to background subtraction. In this case, the component subtracted away from the STM measurement is the vicinal Cu topography. The technique is illustrated in Fig. S2.

Density functional theory (DFT) calculations were performed both at Columbia University and on the computational cluster of the CFN, using the Vienna *Ab-Initio* Simulation Package (VASP) with a plane-wave basis and the projector augmented wave (PAW) method[27, 28]. Different functionals, the local spin density approximation (LSDA), spin polarized generalized gradient approximation (GGA), and the Perdew-Burke-Ernzerhof (PBE) functional, were utilized to asses commonality between their respective results. Calculations typically used 30 or more *k*-points, and periodic slabs were at least 10Å in width and height.

The electronic structure for a system calculated under density functional theory may be deduced if one assumes equivalence with the calculated Kohn-Sham eigenvalues. The band energy diagram and density of states for a ferromagnetic spin phase, computed using a GGA functional, is shown in Fig. S7. Cluster expansion fit parameters were computed using 4 distinct spin configurations of a 4-atom unit cell and fit by a least squares method (Fig. S8).

**Supporting Discussion**

Determining and quantifying the dimerization of the Co chains

Because STM topography measurements have been shown to sometimes be bias dependent, due to tunneling into energy-dependent local density of states (LDOS)[2], the excitation of charge-density modulations[29], or inelastic electron tunneling[30], STM measurements on these chains were repeated at different tip biases ranging broadly from -2V to +2V. These measurements did not show a change in the dimerized appearance of the chains. Another possible source for the non-ideal appearance of the chains is the condition of the tip itself. To rule this out, controlled tip crashes were performed, followed by re-imaging of the chains. Again, no change in the distorted appearance of the chains was found, though a change in the apparent height of the Co atoms was observed. Hence, we conclude that the dimerization instability of the Co chains on the Cu(775) substrate at 5K to be a true structural distortion rather than a bias- or tip-induced electronic effect.

We used STM-derived topography measurements of Co chains to determine the bond-length variation along the chain. The magnitude of the distortion gives insight into the extent and strength of the structural instability and allows for direct comparison to *ab-initio* calculations. An example is shown in Fig. S4. Ideally, the bond-length of two Co atoms constituting the doubled unit cell would be determined by the distance between the local maxima present in the topographical profile irrespective of the profile-line location. However, as can be seen from Fig. 1e, the topography profile of the dimerized Co chain depends on the distance from the step edge. For example, we observe blurring (i.e. coalescing of the Co atom peaks) at the Cu step edge. Similarly, for the case of a single isolated Co dimer, the blurring of the LDOS in a STM topography measurement prevented differentiating the Co atoms that make up the dimer[31]. We therefore employ a more systematic procedure. We present in Fig. S4a line profiles along the wire direction, equidistant from each other and in which the planar cuts are perpendicular to the step terrace, i.e. the (111) plane. The line profiles show a double peak structure that is resolved slightly away from the Cu step edge, as denoted by the length bars. To arrive at the distortion in the Co-Co bond length, the average peak-to-peak (P-P) spacing per profile line was determined and plotted with respect to the average vertical distance of the peaks from the top of the Co wire (the distance from the top of the Co wire was normalized to the apparent Co wire height, which, for this particular example, was approximately equal to the Cu step height of 2.09Å); see Fig. S4b. Note that the distance from the top of the wire is proportional to the distance away from the Cu step edge. Hence, the P-P spacing depends on distance from the step edge. The P-P plot Fig. S4b shows that there is a significant height interval over which a near constant separation between Co atoms within each unit cell is observed. We determine the Co-Co bond length from the spacing in this interval. This means that the bond-length determination is made away from the *Cu* step edge, since in this region the Co atom STM resolution is obscured by coalescing of the Co atom peaks, but not too far from the step edge (where the amplitude of the variation becomes small and the measurement is complicated by other effects), as indicated in Fig S4c. Our measurement shows that the shorter of the two Co-Co bond lengths (i.e. the nearest-neighbor bond length) is 2.0Å (0.1).

Interaction of Co chain with a weakly bound impurity

The interaction of the Co chain with an unidentified impurity was captured by LT-STM, as shown in Fig. S3. $H_2$ and CO are the most common background gases under UHV[32]. Furthermore, it is known that CO preferentially binds to Co but not to Cu[33]. Therefore, we presume the identity of the impurity to be CO.

Co wire STM tip-bias dependent topography present above T$_{Dimerization}$

Besides a lack of uniform nearest-neighbor bond-length, the Co wire phase for T>T$_{Dimerization}$ is distinctly different from that of the dimerization instability phase in that the topography is tip-bias dependent (Fig. S9), indicative of a change in electronic structure. The Co wire nearest-neighbor bond-length modulation is also not necessarily 1-D; at a tip bias of ±500mV, a Co chain exhibits a zig-zag topography. To further compare the modulations of a chain under different tip biases, plots of the profile cut along the chain in Fig. S9 are shown in Fig. S10. Profile line position was taken at peak maximums; unlike the dimerization instability phase, the individual Co atoms do not coalesce together, but rather are resolved at their local maximum. The profiles for occupied and unoccupied states appear relatively similar at a fixed magnitude of tip bias (e.g. ± 100mV, ± 300mV); this is true for a tip bias magnitude of up to at least 500mV. On the other hand, a comparison of line profiles between different tip biases reveals variation in phase shift and amplitude. Due to the finite length of the chain, a Freidel-like oscillation is observed, and has a wavelength of ~25Å. Although T>T$_{Dimerization}$, the chain profile also shows evidence of 2× modulation, particularly at high tip bias and for atoms in the vicinity of the chain's termination.

Orbital polarization of a Co wire in DFT and GGA

We found that DFT using the GGA functional predicts orbital polarization of the $3d_{xy}$ and $3d_{x^2-y^2}$ orbitals (Fig. S7a) as the lowest ground state of the undistorted Co wire. The degree of orbital polarization decreases with increasing distortion (decreasing nearest-neighbor bond length, Fig. S7c). The presence of orbital polarization is the reason for the non-smooth energy phase plots for a Co wire. For completeness, non-orbital polarized energy phase plots are made available in Fig. S5 and S6.

**Supporting References**

**Author Contributions**

N.Z. performed the measurements, analyzed the experimental data, and performed the *ab initio* calculations. C.A.M., A.J.M., N.Z. and J.O. contributed to the theoretical physical understanding of the experiment results. N.Z. and C.A.M. modeled the magnetic coupling. D.P.A., P.Z., and P.S. facilitated low-temperature STM measurements. N.Z., R.M.O., C.A.M., and A.J.M. wrote the paper. P.S., D.P.A., and P.D.J. commented on the manuscript. All authors discussed the results.

S8

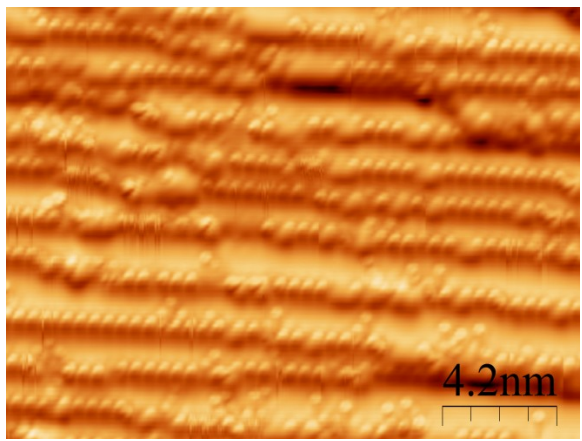

**Figure S1 Co atomic chain self-assembly.** STM image of self-assembled atomic Co chains (visible as linear arrays of approximately spherical dots) on the 14Å average-width terraces of Cu(775) measured at T = 5K. A derivative ($\partial z/\partial y$) filter was applied to accentuate the appearance of the Co decorated step edges. The Co coverage is ~0.14ML. At 5K, the single-atom wide Co chains along the step edge are found to undergo a dimerization distortion along the chain direction. Hence, the sphere like objects are not single Co atoms but rather pairs of Co atoms with separation 22% smaller than the underlying Cu atom spacing. The different appearance of some of the Co atom pairs is due to adsorption of impurity molecules, presumed to be CO (see supporting discussion, Fig. S2). Constant current tunneling parameters: $V_{bias}$ = +2.08V at 9.4nA.

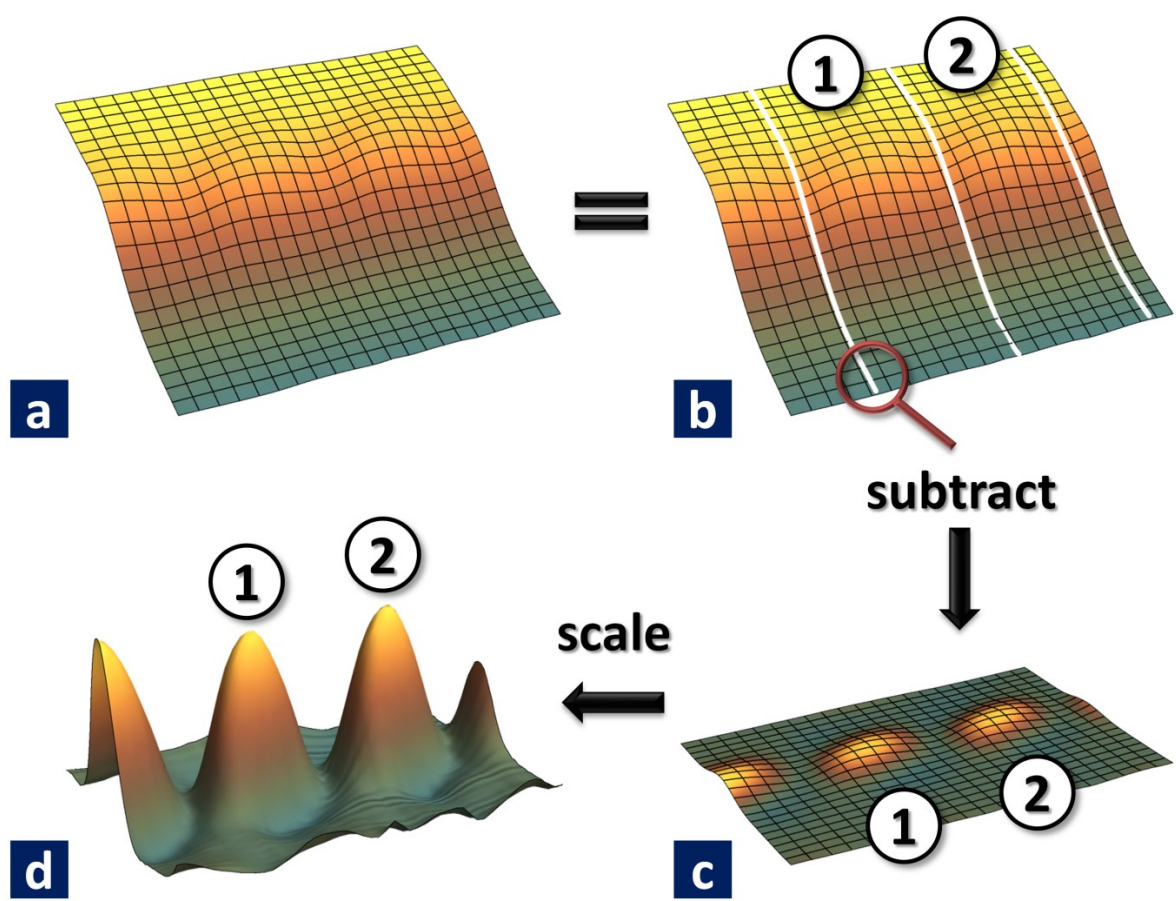

**Figure S2 Delineation of Co atoms by subtraction of Cu step.** (a) STM image of a Co dimerized chain self-assembled at a Cu step edge. The location of the Co atoms can be inferred from the grid lines superimposed on the topography. (b) The same topography as in (a) but with the ×2 distortion boundary denoted by white lines. Within each boundary are a pair of Co atoms; two of these pairs have been marked as 1 and 2. The profiles defined by the white lines approximate the profile of a clean Cu step edge. One of these lines, as indicated by the red circle, has been used for the subsequent subtraction step, which produces the topography shown in (c). (c) Subtracting the aforementioned profile line produces a relatively flat plane with protrusions belonging to the Co-*pair* LDOS. This derived topography can then be vertically scaled to more clearly visualize the Co dimerized chain (d).

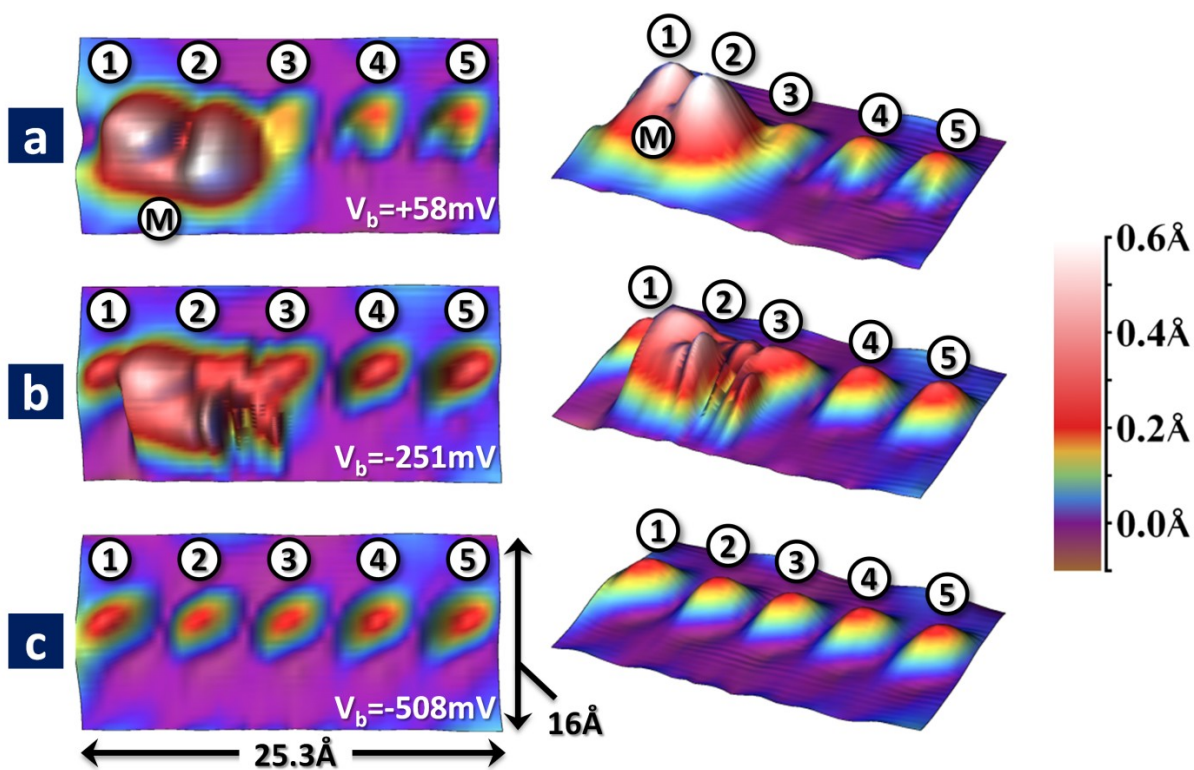

**Figure S3 Interaction of an impurity with a self-assembled Co dimerized chain.** Although the identity of the impurity is unknown, it is guessed to be a CO molecule. The topography images shown here were derived by subtracting the Cu step, as per Fig. S2. (a) An STM topography of a Co dimerized chain, of which 5 Co pairs are visible. In addition, an impurity (labeled M), located in the neighborhood of Co pair 1 and Co pair 2, is also present. The electronic effect of the impurity is to increase the LDOS around Co pairs 1 and 2, as evidenced by the increase in their apparent height. (b) At a higher bias of -0.25V, the tip is able to dislodge the impurity. This is observed in the changing topography, in which scan lines run perpendicular to the chain, from right to left (i.e. from Co-pair 5 to Co-pair 1). After the impurity is dislodged, Co-pair 1 appears with the same contrast as Co pairs 4 and 5. (c) The same chain, after dislodging of the impurity, showing the recovery of the uniform LDOS dimerization for each Co pair. The ability to easily dislodge the impurity using a relatively average tip bias indicates a weakly bound entity to both the Co chain as well as the underlying Cu substrate; in conjunction with the observation of a small apparent volume, these measurements support the identity presupposition of a small molecule such as CO. This observation also shows the relative inactivity (i.e. lack of strong hybridization) of the Co chain with the impurity.

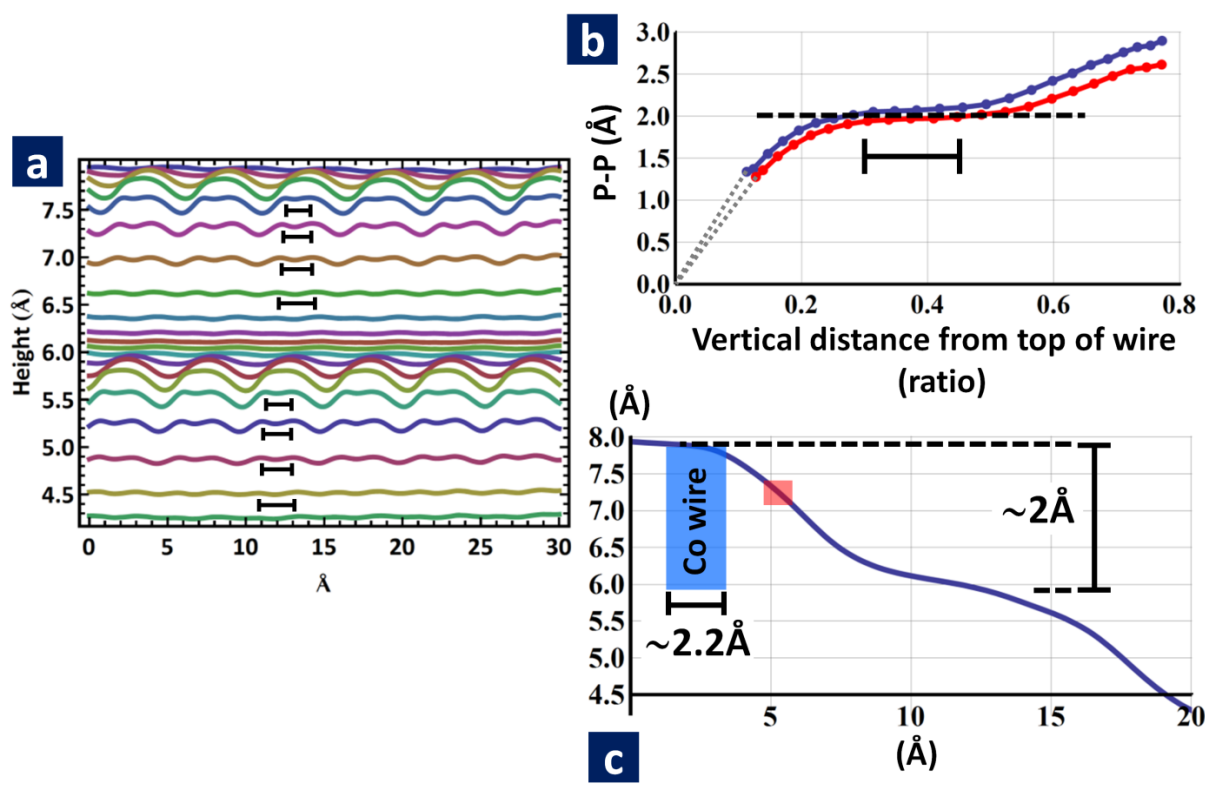

**Figure S4 Measurement and analysis of the Co bond length structural distortion at 5K.** Plots are based on STM measurement shown in Fig. 1e. (a) Height profiles along the Co wire direction, obtained at a series of equally spaced points away from the step edge. Bars are guides to the eye indicating spacing of Co atoms in a dimer. (b) The peak-to-peak (P-P) spacing varies with vertical distance from the top of the Co wire, as shown for the two Co wires by different color curves. The vertical distance from the top of the wire was normalized to the apparent wire height, which, for this particular example, was approximately equal to the Cu step height of 2.09Å. (c) A profile cut perpendicular to the step edge reveals a smooth outline due to the finite radius of the tip and the relatively narrow terrace width. The blue shaded region is a guide to the eye for the approximate position of the Co wire, while the red shaded rectangle corresponds to the region of constant P-P length, denoted in (b) by a bar.

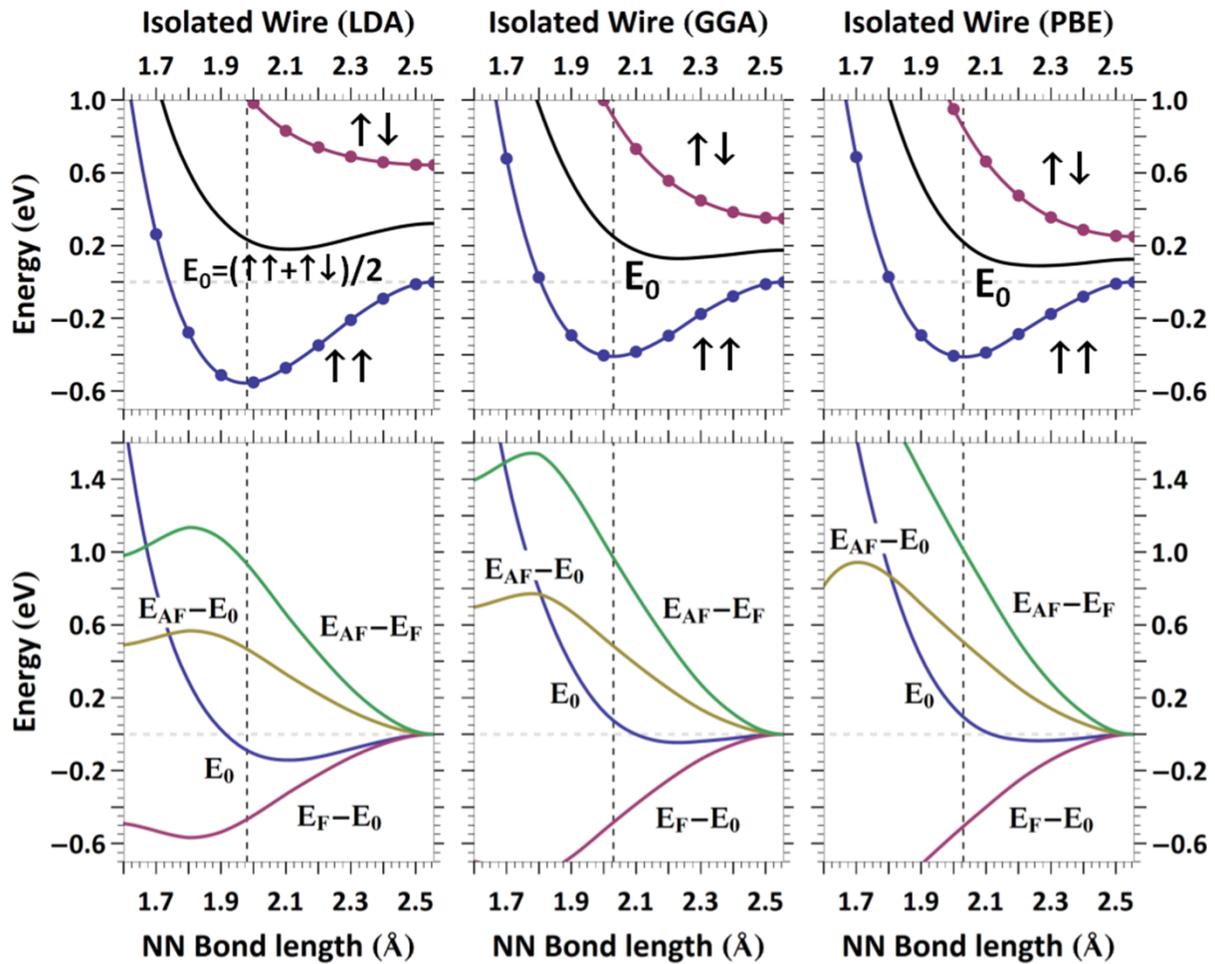

**Figure S5 Co wire distortion phase energies using different DFT functional (*without orbital polarization*).** The top row shows the Co wire energy for different nearest-neighbor distortions; the right vertical axis (2.5561Å) is the case of zero distortion. The energies are referenced to the undistorted ferromagnetic phase. The unit cell size is equal to 2 × Cu atom spacing (2 × 2.5561Å = 5.1122Å). $E_0$ is the average of the ferromagnetic and anti-ferromagnetic phase energies. The points are the actual computed energies, while the smooth lines are 3$^{rd}$-order interpolations. The bottom row shows the change in energy for different combinations of $E_F$, $E_{AF}$, and $E_0$; each combination is referenced to its respective value of the undistorted wire case. The optimal nearest neighbor bond length ($d_{short}$), denoted by the vertical dashed line, is 1.98Å using LDA, and 2.03Å using GGA or PBE.

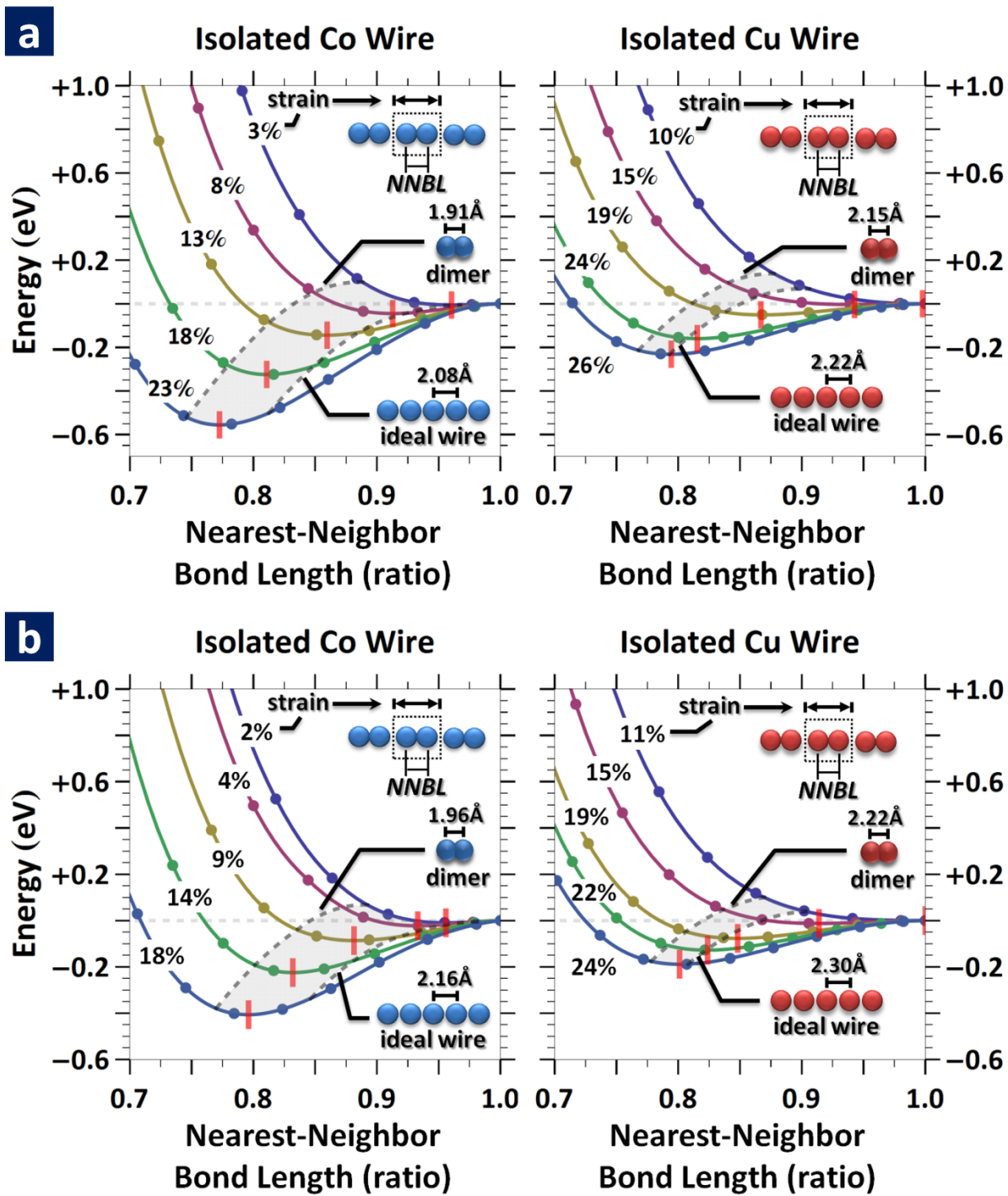

**Figure S6** LDA (a) and non-orbital polarized GGA (b) DFT functional version of Fig. 2b.

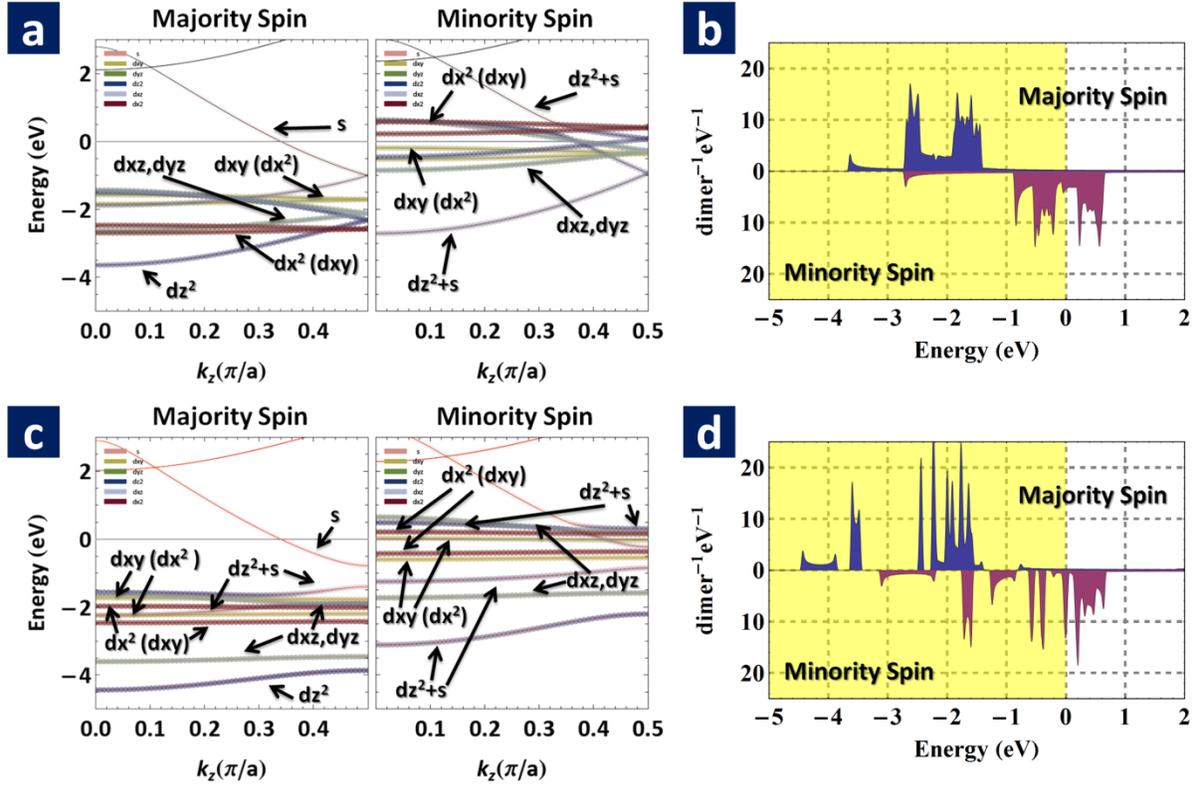

**Figure S7 Electronic structure of isolated Co wires constrained to bulk Cu atom spacing (2.5561Å × 2 unit cell).** Plots (a) and (b) correspond to the non-distorted wire while (c) and (d) correspond to the optimally dimerized case. The z-axis is parallel to the wire. (a), (c) Band energy diagram for majority and minority spin. The $d_{xz}$, $d_{yz}$ orbital derived bands are degenerate, as expected by symmetry. Also, by symmetry, the $d_{z^2}$ and $s$ orbital derived bands are hybridized; the amount of hybridization in each band, however, is not always uniform across a band or across spin populations. The $d_{xy}$, $d_{x^2-y^2}$ orbitals are lifted out of degeneracy due to orbital polarization. Note that overall, the dimerized Co wire is not an insulator; an $s/d_{z^2}$ orbital derived band crosses the Fermi level for both majority and minority spin. (b), (d) The corresponding density of states. Computed using spin polarized GGA.

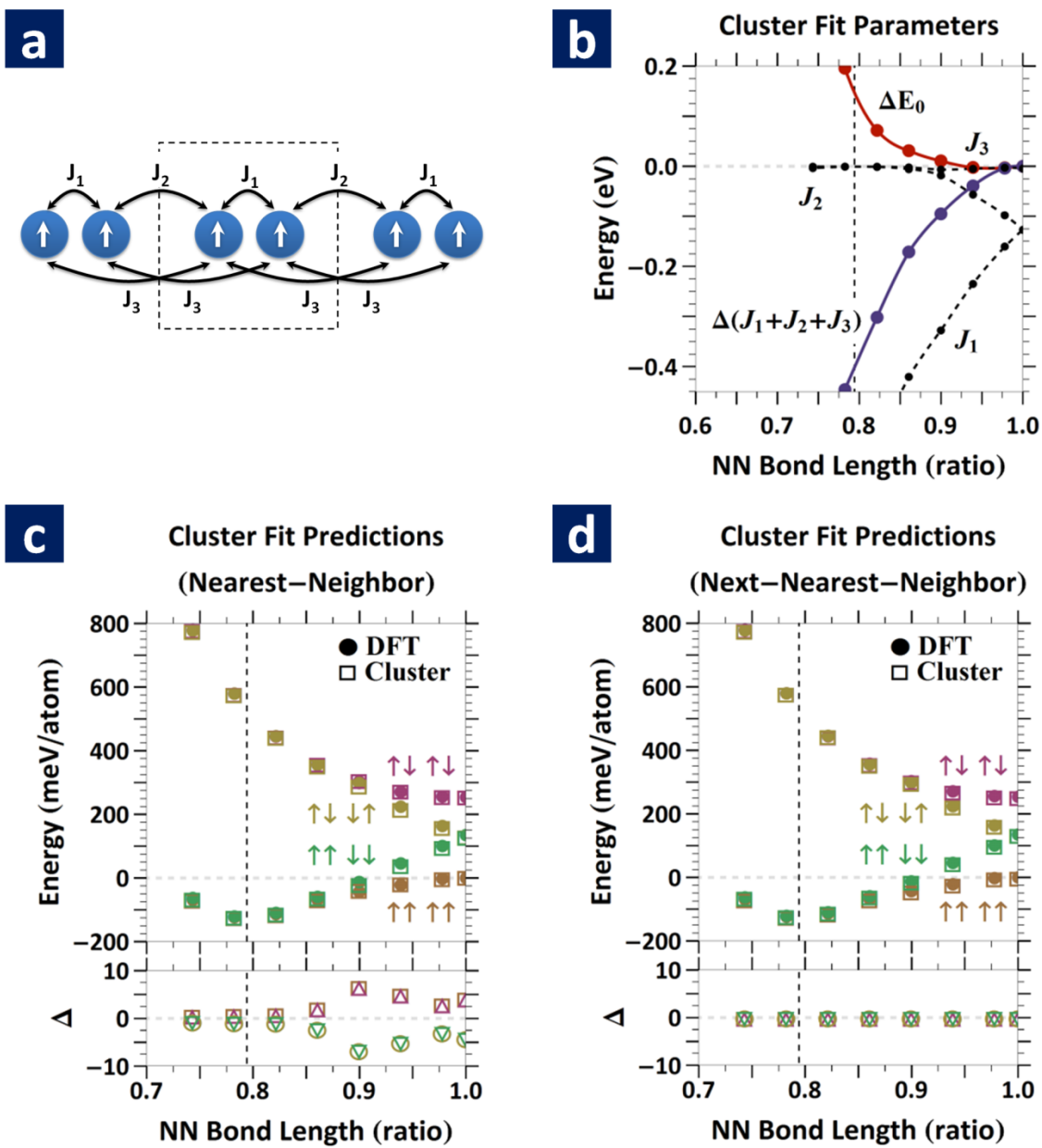

**Figure S8 Cluster fit expansion comparison and errors.** To quantize how well the cluster expansion fits the DFT calculations, we plot the error between DFT and the cluster expansion for both nearest-neighbor ($J_1$ and $J_2$ only) and next-nearest-neighbor ($J_1$, $J_2$ and $J_3$). (a) Illustration of next-nearest-neighbor model. (b) Plot of cluster fit parameters with respect to the short bond length. (c) Energy comparison plot between DFT and nearest-neighbor cluster expansion. The unit cell used for the comparison consisted of 4 atoms, and 4 distinctly different spin configurations were utilized. The error is plotted underneath, with symbol colors corresponding to the respective spin configurations. Note that the error is relatively smaller at the optimal short bond length. (d) Same as (c) but for a next-nearest-neighbor expansion. Energies were computed using spin polarized GGA.

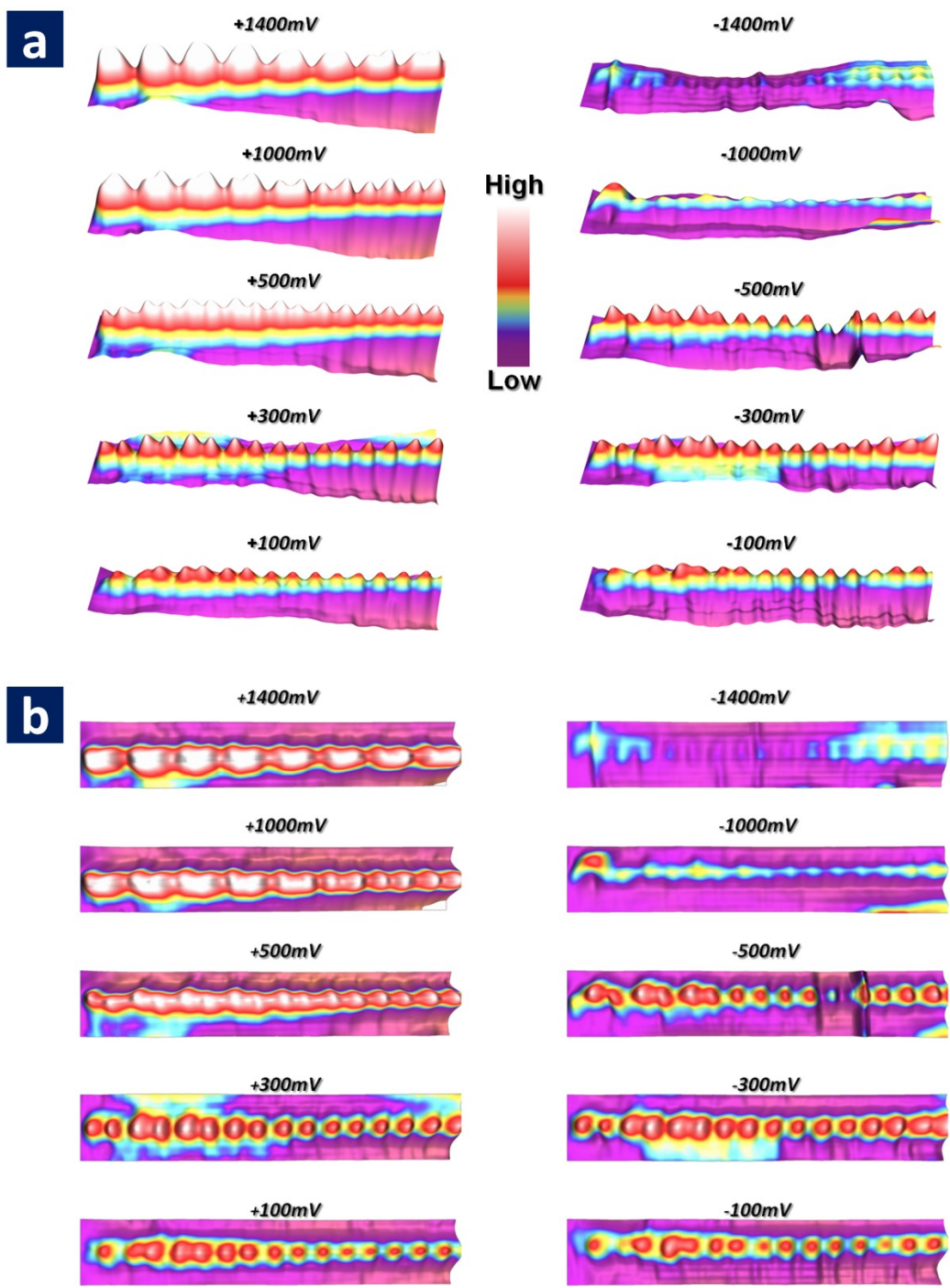

**Figure S9 STM topography of Co wires at 91K.** The technique of substrate subtraction, as described in Fig. S2, was employed in producing these images. This higher temperature phase shows tip-bias dependency, in contrast with the dimerization instability phase present at lower temperatures. (a) and (b) are front and top views, respectively. Note that the chain modulation is not only parallel to the chain, but can appear perpendicular to it; for instance, at ±500mV, the chain appears to have a zig-zag topography. The discontinuity in the topography at −500mV is due to a tip change.

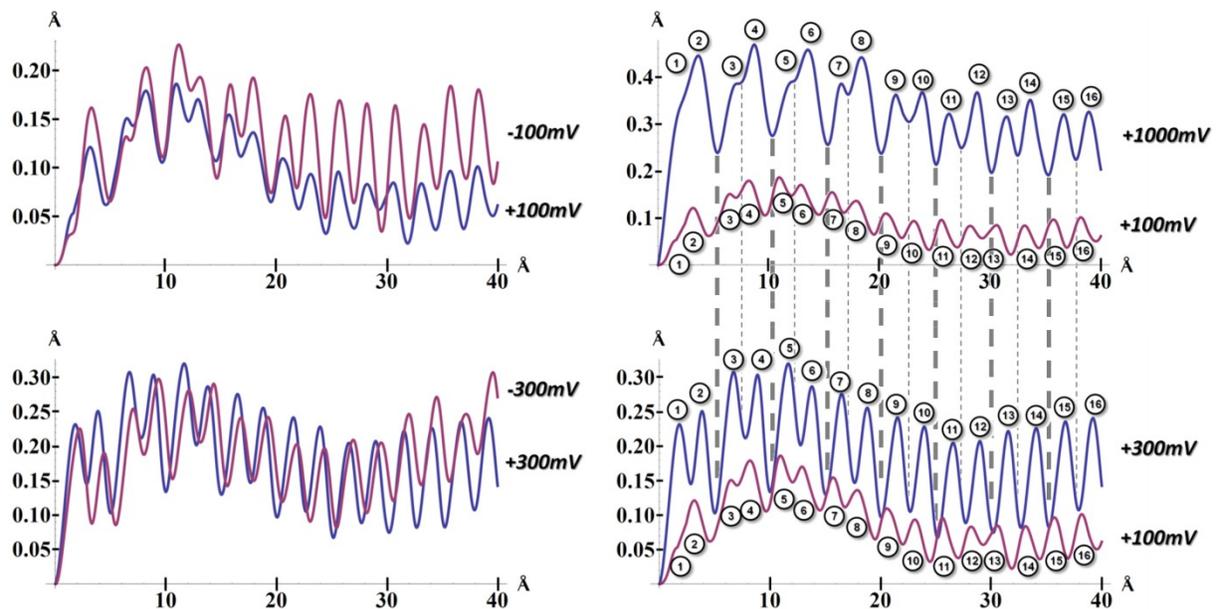

**Figure S10 Line profiles along Co chain of Fig. S9.** Profile line position was taken at peak maximums; unlike the dimerization instability phase, the individual Co atoms do not coalesce together, but rather are resolved at their local maximum. The profiles for occupied and unoccupied states appear relatively similar at a fixed magnitude of tip bias (e.g. ± 100mV, ± 300mV); this is true for a tip bias magnitude of up to at least 500mV. On the other hand, a comparison of line profiles between different tip biases reveals variation in phase shift and amplitude. The numbered peaks and dashed vertical lines serve as a guide for the eye. The chain terminates on the left (0 Å); a Friedel-like oscillation is observed due to the finite length of the chain, and has a wavelength of ~25Å.

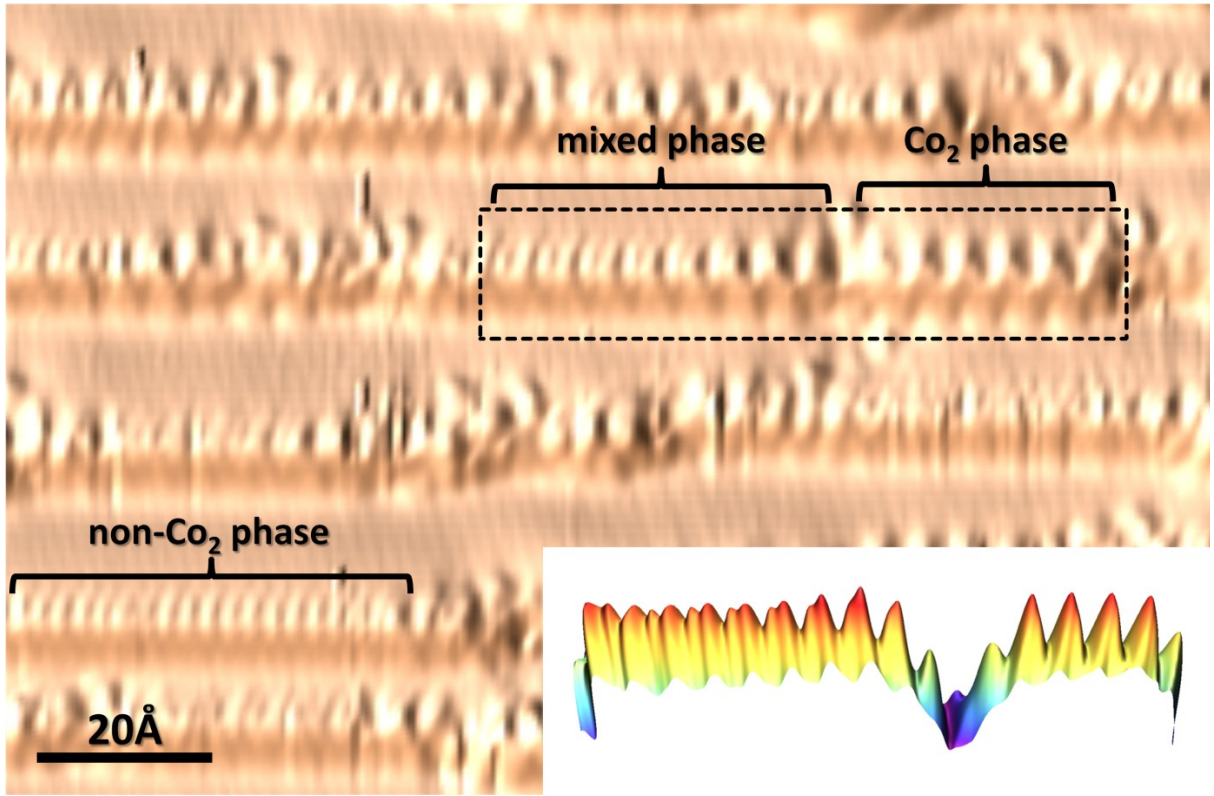

**Figure S11 Co-existence of two phases of a Co chain at T~$T_{Dimerization}$ = 81K.** STM topography was measured with tip bias of +238mV. Three chains are delineated. A chain with a dimerization instability (labeled "$Co_2$ phase") exhibits a clear 2× distortion as well as coalescing of the LDOS peaks. A chain without a dimerization instability (labeled "non-$Co_2$ phase") does not show a 2× distortion; instead, the constituent individual Co atoms are clearly delineable and the nearest neighbor bond length is closer to that of the underlying Cu substrate; the topography is similar to that at T>$T_{Dimerization}$ (see Fig. S6). The chain labeled "mixed phase" exhibits regions with varying similarity to the dimerization instability phase. In comparison to measurements at this temperature, STM measurements taken at T>$T_{Dimerization}$ exhibit a uniform phase topography. The inset is a rendering of the topography enclosed in the dashed rectangle after applying the substrate subtraction technique of Fig. S2.